\documentclass[preprint]{aastex}
\newcounter{species}
\def\ion#1#2{\setcounter{species}{#2}#1$\;${\scriptsize\Roman{species}}\relax}

\shorttitle{$H$-band M Dwarf Metallicities}
\shortauthors{Terrien et al.}

\begin{document}

\title{An $H$-band Spectroscopic Metallicity Calibration for M Dwarfs}

\author{Ryan C. Terrien\altaffilmark{1,2}, Suvrath Mahadevan\altaffilmark{1}, Chad F. Bender\altaffilmark{1}, Rohit Deshpande\altaffilmark{1}, \\ Lawrence W. Ramsey\altaffilmark{1}, and John J. Bochanski}
\affil{Department of Astronomy \& Astrophysics, The Pennsylvania State University, University Park, PA, 16802, USA}

\altaffiltext{1}{Center for Exoplanets and Habitable Worlds, The Pennsylvania State University, University Park, PA, 16802, USA}
\altaffiltext{2}{rct151@psu.edu}

\begin{abstract}
We present an empirical near-infrared (NIR) spectroscopic method for estimating M dwarf metallicities, based on features in the $H$-band, as well as an implementation of a similar published method in the $K$-band. We obtained {\it R}$\sim$2000 NIR spectra of a sample of M dwarfs using the NASA IRTF-SpeX spectrograph, including 22 M dwarf metallicity calibration targets that have FGK companions with known metallicities. The $H$-band and $K$-band calibrations provide equivalent fits to the metallicities of these binaries, with an accuracy of $\pm$0.12 dex. We derive the first empirically calibrated spectroscopic metallicity estimate for the giant planet-hosting M dwarf GJ 317, confirming its super-solar metallicity. Combining this result with observations of eight other M dwarf planet hosts, we find that M dwarfs with giant planets are preferentially metal-rich compared to those that host less massive planets. Our $H$-band calibration relies on strongly metallicity-dependent features in the $H$-band, which will be useful in compositional studies using mid to high resolution NIR M dwarf spectra, such as those produced by multiplexed surveys like SDSS-III APOGEE. These results will also be immediately useful for ongoing spectroscopic surveys of M dwarfs.
\end{abstract}


\section{Introduction}
Chemical analysis of M dwarfs is notoriously difficult due to the complex molecular spectra of their cool atmospheres \citep{1976A&A....48..443M,1989ARA&A..27..701G}. Substantial progress in metallicity estimation for M dwarfs has been made recently, both through photometry and spectroscopy, and motivated in large part by by the question of whether the metallicity-giant planet (\mbox{$>0.5 M_{\mathrm{Jupiter}}$}) correlation \citep{1997MNRAS.285..403G,2004A&A...415.1153S,2005ApJ...622.1102F} extends to low-mass stars.

Photometric techniques are based on the higher optical opacity in metal-rich stars, which shifts visible light into the infrared \citep[e.g.][]{Delfosse:2000ur}. \cite{2005A&A...442..635B} first calibrated this relationship, and later groups have modified it and addressed its limitations \citep{2009ApJ...699..933J,2010A&A...519A.105S}. \cite{2011arXiv1110.2694N} evaluated the published photometric calibrations, and found \cite{2010A&A...519A.105S} to have the lowest dispersion (rms $=0.19\pm0.03$ dex). Photometric techniques have generally been limited by requiring absolute magnitudes of the targets, although the new technique presented in \cite{Johnson:2011wq} depends on color alone.

Empirical spectroscopic techniques have also been developed, which rely on calibrations using optical \citep{2006PASP..118..218W} or near-infrared (NIR) $K$-band \citep[][hereafter, RA10]{RojasAyala:2010ht} feature strengths. RA10 used \ion{Na}{1}, \ion{Ca}{1}, and H$_{2}$O features to model metallicity, and calibrated their model using M dwarf companions to FGK stars, with an uncertainty of $\pm0.15$ dex.\footnote[3]{At submission of this paper, \citeauthor{RojasAyala:2010ht} have submitted an updated calibration \citep{RojasAyala:2011te} with an uncertainty of 0.14 dex.} Direct spectroscopic modeling has also been attempted on a small number of stars \citep[][hereafter O11]{Bean:2006el,2007PASP..119...90M,Onehag:2011to}, but is still prohibitively difficult to be used on a large scale due to the intrinsic faintness of M dwarfs, their dense spectra, and incomplete linelists.

We are developing a target list in support of the Habitable Zone Planet Finder \citep{2010SPIE.7735E.227M}, a NIR precision radial velocity spectrograph for finding planets around M dwarfs. Metallicity estimates will be an important parameter of this target list, so that we may sample a wide range of stellar metallicities. Our overall list is drawn from the all-sky nearby M dwarf catalog presented in \cite{2011arXiv1108.2719L}, and we are carrying out a campaign on the NASA Infrared Telescope Facility (IRTF) to obtain {\it R}$\sim$2000 spectra of these stars  and estimate their metallicities using the method described herein. M dwarfs will also be targeted in high-resolution surveys, such as the Sloan Digital Sky Survey III \citep{2011AJ....142...72E} APOGEE survey \citep{2008AN....329.1018A}, which will deliver {\it R}$\sim$20,000 $H$-band spectra of hundreds of M dwarfs at multiple epochs. 

Here, we report on a new NIR spectroscopic calibration of metallicity-sensitive features in the $H$-band, calibrated with well-separated M dwarf companions to FGK stars with known metallicities. This calibration yields a smaller scatter than the RA10 $K$-band calibration. The main promise of the $H$-band technique is that the relevant metallicity-sensitive features will be included in APOGEE high-resolution $H$-band spectra of M dwarfs, allowing a high-resolution metallicity calibration and potentially direct spectroscopic modeling like that accomplished by O11. In \S2, we describe our observations, in \S3, we explain our analysis and error estimates, and in \S4, we discuss our calibrations and our analysis of M dwarf planet host metallicities, which indicates that M dwarfs with giant planets are preferentially metal-rich compared to those that host less massive planets.

\section{Observations}
During August and November 2011, we observed more than 200 M dwarfs using the SpeX spectrograph \citep{2003PASP..115..362R} on the NASA IRTF. For calibration of our empirical metallicity scale, our target list included 22 M dwarfs in binary systems with FGK primaries that have [Fe/H] measurements from SPOCS \citep{2005ApJS..159..141V}, which adopts an uncertainty in [Fe/H] of $\pm$0.030 dex. The properties of these systems, which include two new proper motion-selected binary systems \citep[from][]{2011arXiv1108.2719L}, are listed in Table \ref{mcal_table}. We are currently analyzing the remainder of our M dwarf sample, and will publish these results in a separate paper.

We operated SpeX in the short cross-dispersed (SXD) mode with the 0.3x15" slit (2 pixels per resolution element), which yielded {\it R}$\sim$2000 spectra spanning 0.8-2.4 $\mu$m. SpeX uses a standard nodding technique for sky subtraction. For our standard targets, we reached a minimum S/N$\approx$150 per pixel, while for the calibrators, we obtained S/N$\approx$200 per pixel or greater. In clear conditions, we were able to achieve S/N$\approx$150 on an $H$=9 target with 10 minutes of integration.

We extracted our spectra using the SpeXTool package \citep{2004PASP..116..362C}, which automatically subtracts the AB nods, divides out flat-field exposures, wavelength calibrates, and performs optimal extraction. To achieve the best data quality, and avoid known instrument shifts with large telescope slews, we obtained flat-field exposures and argon lamp calibration sequences throughout each night. Each spectrum was telluric-corrected and flux-calibrated using a nearby A0V star, or a close spectral type if there was no nearby A0V star. Our telluric standards were always within 0.1 airmass of the targets, and usually much closer. They were observed within 30-60 minutes of the targets. Telluric correction and flux calibration were performed with the \textit{xtellcor} telluric correction code \citep{2003PASP..115..389V}.

\section{Analysis and Results}
\begin{figure*}
\begin{center}
\rotate{}
\includegraphics[scale=.7]{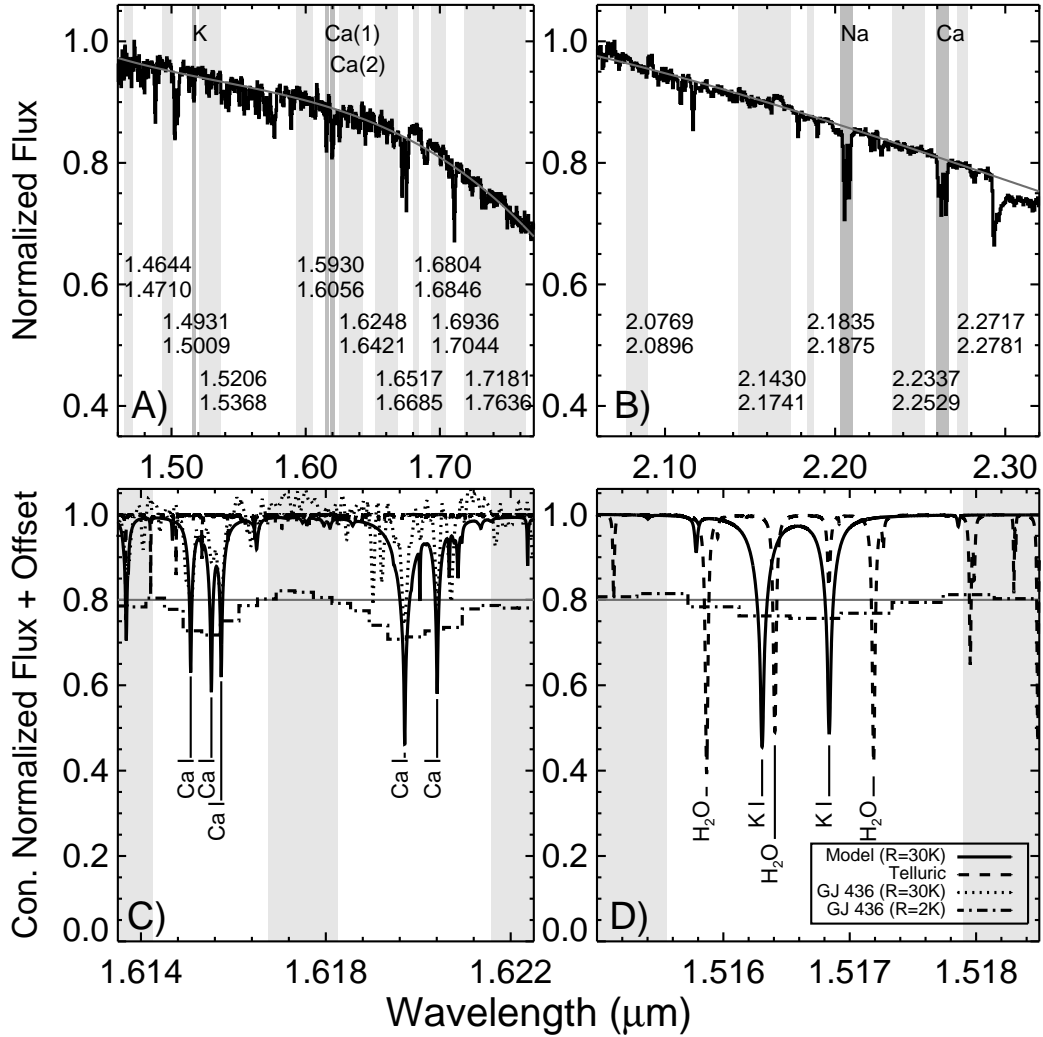}
\caption{(A,B) The $H$ and $K$-band orders of GJ 436, with continuum regions marked in light gray, feature regions marked in dark gray, and the continuum fit overlaid. The left and right limits of the unshifted continuum regions are noted in $\mu$m, and the spectra have been normalized by a constant. (C,D) High-resolution, continuum-normalized spectra of the $H$-band \ion{Ca}{1} and \ion{K}{1} features in GJ 436. Overlaid are: a model spectrum produced using linelists from VALD \citep{1995A&AS..112..525P}, updated log($gf$) values from \cite{1999ApJS..124..527M}, and the SYNTH3 code \citep{2007pms..conf..109K}, and convolved to {\it R}$\sim$30,000; an observed spectrum of GJ 436 at {\it R}$\sim$30,000 \citep{2005AJ....129..402B}; a high resolution telluric model spectrum \citep{2005JQSRT..91..233C} that is dominated by narrow CO lines in the \ion{Ca}{1} feature and by H$_{2}$O in the \ion{K}{1} feature; and our IRTF observations of GJ 436 ({\it R}$\sim$2000) with the continuum shown and offset by 0.2. The white regions show the feature windows.
\label{cafeature}}
\end{center}
\end{figure*}

Our metallicity calibration is based on features in the $H$ and $K$-bands. While our $K$-band calibration is similar to RA10, we use a different flux windows and continuum estimation regions, as described below. We also measured the equivalent widths (EWs) of two $H$-band features, which we label the \ion{K}{1} (1.52$\mu$m) and the \ion{Ca}{1} features (1.62$\mu$m) based on their dominant elemental opacities. The decomposition of the $H$-band features is shown in Figure~\ref{cafeature}. The \ion{K}{1} feature is weak and contains a telluric H$_{2}$O feature. The $H$-band \ion{Ca}{1} feature contains multiple atomic lines and is composed of the sum of two individual features, hereafter denoted Ca(1) and Ca(2). In order to perform our $H$-band analysis, we required S/N$\approx$150 per pixel, compared to the S/N$\approx$100 for the stronger $K$-band features. 

\subsection{Equivalent Width Calculation}
The first step in measuring the EWs was to define a standard width and location for each feature. The width had to be sufficient to contain the feature at all strengths but not so wide as to contain nearby features. Our fixed feature widths, $W$, were:
\begin{eqnarray*}
W_{\mathrm{Na,K-band}} & = & 71 \,\mathrm{ \AA}, \\
W_{\mathrm{Ca,K-band}} & = & 74 \,\mathrm{ \AA}, \\
W_{\mathrm{Ca(1),H-band}} & = & 25 \,\mathrm{ \AA}, \\
W_{\mathrm{Ca(2),H-band}} & = & 33 \,\mathrm{ \AA}, \\
W_{\mathrm{K,H-band}} & = & 24 \,\mathrm{ \AA}. \\
\end{eqnarray*}
As discussed in \S\ref{errorsection}, the feature widths were chosen to produce the best agreement between our analytic and Monte Carlo error estimates, signifying that the EW measurements were robust to small changes in the spectrum. These windows were positioned on centers $C$:
\begin{eqnarray*}
C_{\mathrm{Na,K-band}} & = & 2.2074 \,\mu\mathrm{m}, \\
C_{\mathrm{Ca,K-band}} & = & 2.2638 \,\mu\mathrm{m}, \\
C_{\mathrm{Ca(1),H-band}} & = & 1.6159 \,\mu\mathrm{m}, \\
C_{\mathrm{Ca(2),H-band}} & = & 1.6203 \,\mu\mathrm{m}, \\
C_{\mathrm{K,H-band}} & = & 1.5171 \,\mu\mathrm{m}. \\
\end{eqnarray*}

Calculation of the EWs also requires an estimate of the pseudo-continuum for normalization. To obtain this, we selected regions in each order that bracketed our features and contained no strong features themselves. The continuum regions are listed in Figure \ref{cafeature}. These regions were fit by a fourth-order Legendre polynomial using the IDL \textit{svdfit} routine, and the fits were visually confirmed.

We then accounted for radial velocity (RV), opting to use the strong features provided by the entire $K$-band order and not the comparatively noisy $H$-band. As a template, we used the high-S/N spectrum of the M1.5V star HD36395 from the IRTF spectral library \citep{2009ApJS..185..289R}. A fit to the cross-correlation peak yielded a velocity offset, which was applied to all the feature and continuum regions for each target.

EW measurement for each feature then proceeded in the standard fashion:
\begin{equation}
\mathrm{EW}_{\lambda} = \sum_{i} \left[ 1 - \frac{I(\lambda_{i})}{I_{c}(\lambda_{i})} \right]\,\Delta\lambda_{i},
\end{equation}
where the feature spans pixels $i$, $\lambda_{i}$ is the wavelength at pixel $i$, $I$ is the intensity, and $I_{c}$ is the psuedo-continuum intenstiy. We included in this sum fractional ``edge" pixels which fell inside the feature windows, weighted by the proportion of the pixel contained in the window. For the H-band \ion{Ca}{1} feature, we combined the two components:
\begin{equation}
\mathrm{EW}_{\mathrm{Ca,H-band}}=\mathrm{EW}_{\mathrm{Ca(1),H-band}}+\mathrm{EW}_{\mathrm{Ca(2),H-band}}.
\end{equation}

As suggested by RA10, we accounted for temperature effects by including in our model H$_{2}$O indicies in both the $K$ and $H$-bands, as the strength of H$_{2}$O absorption in these bands tracks spectral type \citep{2010ApJ...722..971C}. These indicies are defined as:
\begin{eqnarray*}
\mathrm{H}_{2}\mathrm{O\textendash K}\hspace{-.3cm}&=&\hspace{-.3cm}\frac{ \langle \mathcal{F}(2.180 - 2.200) \rangle / \langle \mathcal{F}(2.270-2.290) \rangle} { \langle \mathcal{F}(2.270-2.290) \rangle / \langle \mathcal{F}(2.360-2.380) \rangle}, \\
\mathrm{H}_{2}\mathrm{O\textendash H}\hspace{-.3cm}&=&\hspace{-.3cm}\frac{ \langle \mathcal{F}(1.595-1.615) \rangle / \langle \mathcal{F}(1.680-1.700) \rangle} { \langle \mathcal{F}(1.680 - 1.700) \rangle / \langle \mathcal{F}(1.760 - 1.780) \rangle},
\end{eqnarray*}
where $\langle \mathcal{F}(a-b) \rangle$ is the mean flux in the range bounded by $a$ and $b$ in $\mu$m. A calibration of H$_{2}$O-K based on the stars in the IRTF Spectral Library \citep{2009ApJS..185..289R} provides the spectral type estimates listed in Table \ref{mcal_table}.

\subsection{Equivalent Width Errors\label{errorsection}}
We used two independent methods to estimate uncertainties in our EW measurements. The first was an analytic formalism described in \cite{1992ApJS...83..147S}. Briefly, this method accounts both for errors in intensity and errors in the continuum fit, using the measured uncertainty from the extraction and the covariance matrix from the least squares fit to the pseudo-continuum. This yielded average relative errors for our EWs of 1-3\%.

Since this method did not account for errors induced by our RV correction, we also computed Monte Carlo errors. We performed 100 trials for each target, jittering each pixel by a draw from a normal distribution defined by its extracted error ($\sim$0.3-0.5\%), and repeated the entire EW analysis on each iteration. The spread in the recovered EW values included errors in RV shifts, intensity, and pseudo-continuum fitting. In Table \ref{mcal_table}, we report the more conservative Monte Carlo uncertainties. We selected our feature definitions by minimizing (through trial-and-error) the discrepancies between our analytic and Monte Carlo errors. For all except two metallicity calibrators, the Monte Carlo and analytic error estimates were consistent, their uncertainties being dominated by intensity and continuum-fitting uncertainties rather than RV uncertainties.

We note that some of our observations were performed in conditions with highly variable cloud cover and water column. We diagnosed this issue by reducing individual frames of a selection of targets which had high S/N in each frame. In poor conditions, we found that the EW of the weak $H$-band \ion{K}{1} feature could vary between frames by as much as 20-30\%($\sim$0.2\AA). We suspect this is due to strong nearby H$_{2}$O features (as shown in Figure \ref{cafeature}). Simulating strong H$_{2}$O features \citep[][]{2005JQSRT..91..233C} in varying water column and calculating the resultant EWs showed that a realistic variation of 1mm of water column yielded $\Delta$EW$\approx$0.2\AA, enough to account for the observed fluctuations.

\subsection{Correlations}
\begin{figure*}
\begin{center}
\includegraphics[scale=.7]{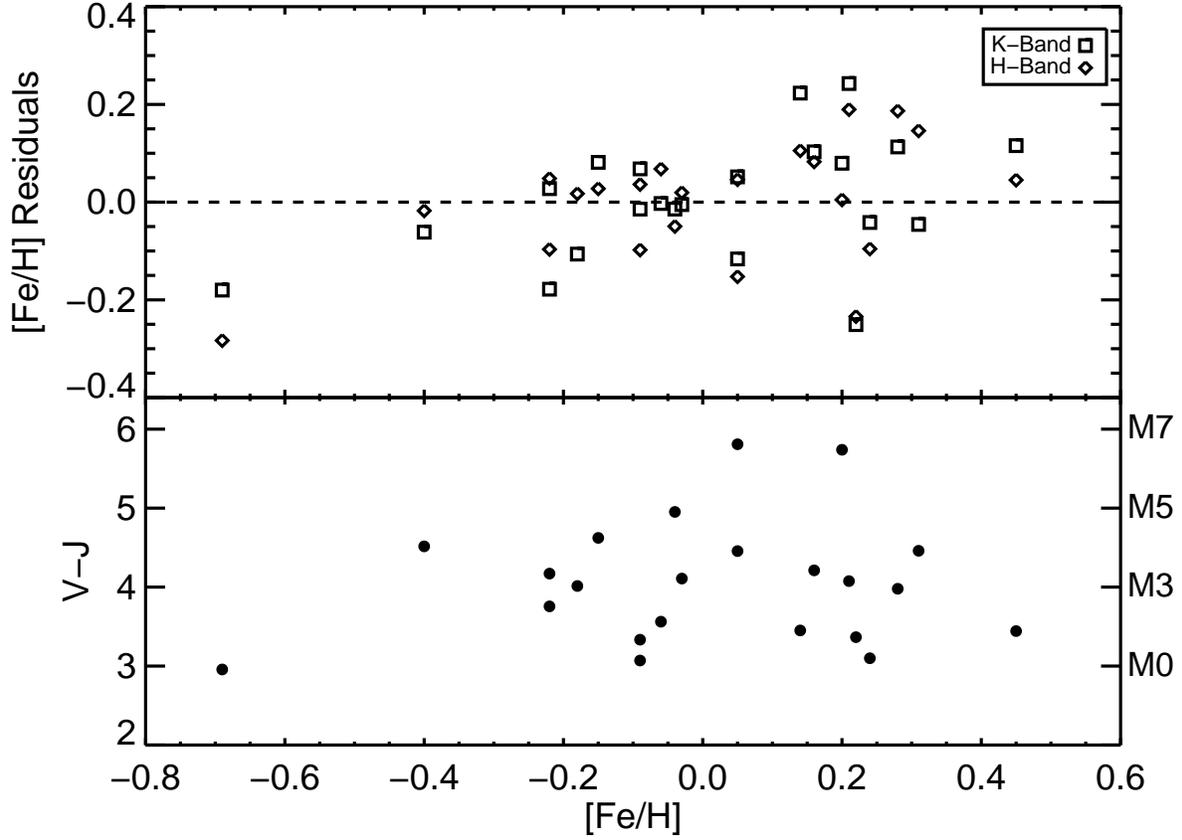}
\caption{Top: The residuals between the K/H band metallicity estimates and the SPOCS  \citep{2005ApJS..159..141V} primary metallicity. Bottom: The distribution of calibrators in color and metallicity, using visual and 2MASS J magnitude \citep{Skrutskie:2006hl}, showing the well-constrained regions for the calibrations: approximately $-0.25<\mathrm{[Fe/H]}<0.3$ and M0 to M5. \label{absmetal}}
\end{center}
\end{figure*}

We performed two least-squares linear regressions on our calibration targets, with metallicity as the response: one with the $K$-band features and the H$_{2}$O-K index, and one with the $H$-band features and the H$_{2}$O-H index. For the $K$-band, we found the best fit equation
\begin{eqnarray*}
[\mathrm{Fe}/\mathrm{H}]_{\mathrm{K-band}}&=&0.132(\mathrm{EW}_{\mathrm{Na}}) + 0.083(\mathrm{EW}_{\mathrm{Ca}})\\&& - 0.403(\mathrm{H}_{2}\mathrm{O\hspace{-1mm}-\hspace{-1mm}K}) - 0.616, \\
\sigma([\mathrm{Fe}/\mathrm{H}]_{\mathrm{K-band}}) &=& 0.12. \\
\end{eqnarray*}
For the $H$-band, we found:
\begin{eqnarray*}
[\mathrm{Fe}/\mathrm{H}]_{\mathrm{H-band}} &=& 0.340(\mathrm{EW}_{\mathrm{Ca}}) + 0.407(\mathrm{EW}_{\mathrm{K}})\\&& + 0.436(\mathrm{H}_{2}\mathrm{O\hspace{-1mm}-\hspace{-1mm}H})-1.485, \\
\sigma([\mathrm{Fe}/\mathrm{H}]_{\mathrm{H-band}}) &=& 0.12, \\
\end{eqnarray*}
where $\sigma$ is the dispersion of the residuals, and an average of the $K$ and $H$-band metallicity estimates also yields a residual dispersion of $\sigma([\mathrm{Fe}/\mathrm{H}]_{\mathrm{avg}}) = 0.12$. Residuals to these fits are shown in Figure~\ref{absmetal}. The observed fluctuations of the weak \ion{K}{1} feature suggest that the $H$-band metallicity estimate is more susceptible to atmospheric variability. For example, a variation of 0.2\AA~in the H$_{2}$O feature would yield $\Delta$[Fe/H]$\approx$0.08 dex.

For comparison with other empirical methods, we present the residual mean squares ($RMS_{\mathrm{p}}$) and the adjusted squared multiple correlation coefficients $R^{2}_{a}$ for our models and others \cite[from RA10 and][]{2010A&A...519A.105S}. Our $H$ and $K$-band models provide equivalent fits: both have $RMS_{\mathrm{p}}=0.02$ and they have $R^{2}_{a}=0.74$ and $R^{2}_{a}=0.73$, respectively. Our results are an improvement over those presented by \cite{RojasAyala:2011te} ($RMS_{\mathrm{p}}=0.02$, $R^{2}_{a}=0.67$), RA10 ($RMS_{\mathrm{p}}=0.02$, $R^{2}_{a}=0.63$), JA09 ($RMS_{\mathrm{p}}=0.04$, $R^{2}_{a}=0.059$), SL10 ($RMS=0.02$, $R^{2}_{a}=0.49$), and \cite{2011arXiv1110.2694N} ($RMS_{\mathrm{p}}=0.03$, $R^{2}_{a}=0.43$). Each quantity is reported for the calibration set with which it was created. 

As shown in Figure \ref{absmetal}, our calibration is well-constrained for approximately $-0.25<\mathrm{[Fe/H]}<0.3$. It is poorly constrained for late ($>$M5), low-metallicity M dwarfs. We note that increasing the order of our model (e.g.\ to quadratic) does improve our fit, but this is mostly due to the single point with [Fe/H]$_{\mathrm{SPOCS}}=-0.69$, for which our calibrations overestimate the metallicity compared to the primary star (as does RA10).


\section{Discussion}
\begin{figure*}
\begin{center}
\includegraphics[scale=.7]{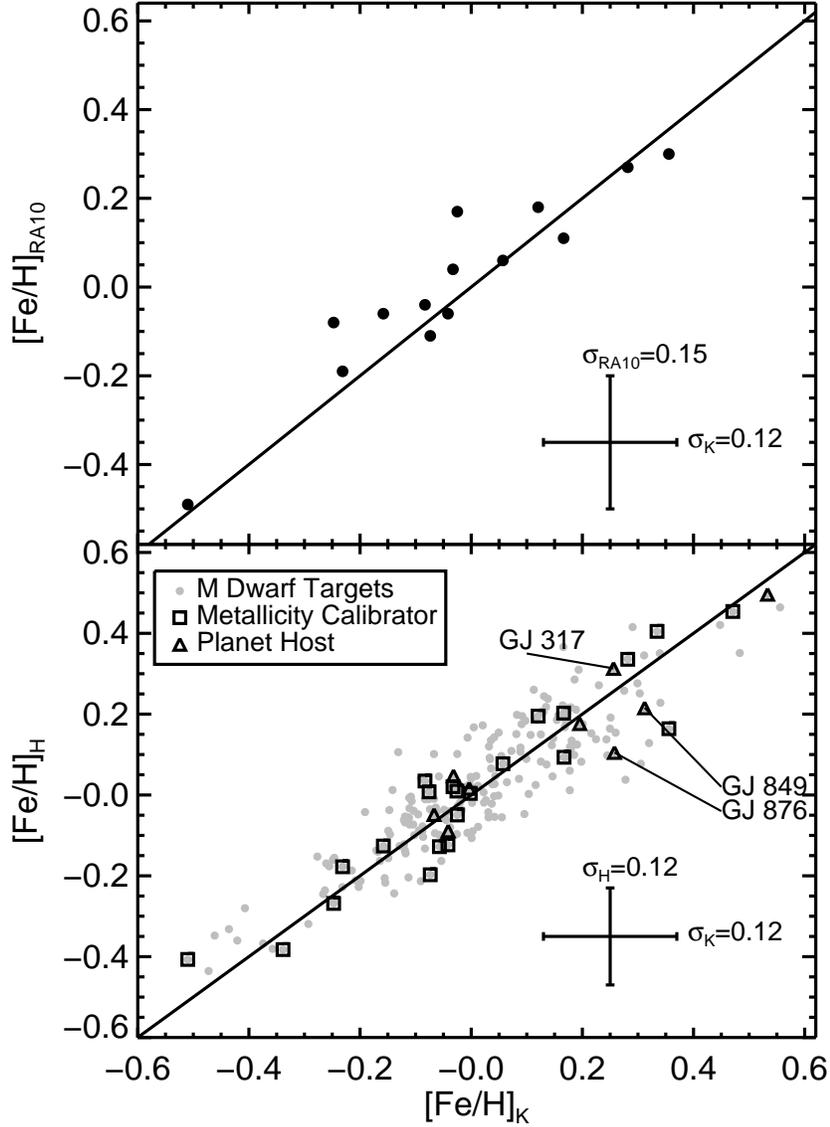}
\caption{Top: Our $K$-band metallicity estimates compared to the RA10 estimates for the metallicity calibrators we have in common. In both panels, the line represents equivalence between the two estimates. The results are consistent. Bottom: A comparison of the $K$ and $H$-band metallicity estimates for all program stars. The $H$-band estimates track the $K$-band estimates well. GJ 317, GJ 849, and GJ 876 are marked.\label{kvsh}}
\end{center}
\end{figure*}

Figure~\ref{kvsh} shows our calibration applied to our entire M dwarf sample, and demonstrates the equivalence of our $K$ and $H$-band metallicity estimates, as well as the consistency of our implementation of the RA10 $K$-band calibration. We suspect that some of the scatter between the $K$ and $H$-band estimates is produced by the atmospheric variations discussed in \S\ref{errorsection}. Other sources of scatter may include variations in stellar surface gravity, magnetic fields, and activity. Although progress is being made in characterizing activity-sensitive NIR features in M dwarfs \citep[e.g.][]{2012ApJ...745...14S}, our features have not been examined in this context. However, we did estimate the gravity sensitivity of our features using the BT-Settl \citep{Allard:2010wp} models. We applied our EW algorithm to stellar models with T=3100K, solar metallicity, and log($g$) = 4.0, 4.5, and 5.0. This showed that the Ca features were not strongly gravity-sensitive, but that an increase of 0.1 in log($g$) yielded $\Delta$EW$_{\mathrm{Na}}$$\approx$0.2\AA~and $\Delta$EW$_{\mathrm{K}}$$\approx$0.1\AA. These changes translate to $\Delta$[Fe/H]$\approx$0.03 and 0.04 dex, respectively.

\subsection{Calibration}
The main disadvantage of the $H$-band method is that the $H$-band features are much weaker than those in the $K$-band, and therefore higher S/N spectra and non-variable conditions are required in order to accurately measure the feature strengths. However, the $H$-band offers three important advantages:
\begin{itemize}
\item The $H$-band calibration allows an independent metallicity estimate from the $K$-band calibration, in a spectral region less plagued by thermal noise. The information content of the $H$-band may be useful to other groups which are pursuing similar programs to ours.
\item The $H$-band relation is superior to previous models in terms of $R^{2}_{a}$. That this relation performs as well as it does despite its sensitivity to atmospheric effects is quite promising. Moreover, higher resolution observations can mitigate much of the atmospheric effects with detailed telluric modeling.
\item The $H$-band model locates features that are strongly correlated with metallicity. These features and our calibration can be used as a starting point for M dwarf metallicity studies with high-resolution surveys like APOGEE. A recent analysis by O11 demonstrates that high resolution spectroscopic modeling of early M dwarfs is feasible in the $J$-band, raising the possibility that the $H$-band might also be amenable to such techniques. 
\end{itemize}

\subsection{Metallicity of Planet Hosts}
We have also applied our calibration to several planet hosts, listed in Table \ref{planettable}. Our measurement of GJ 317 represents the first empirically calibrated spectroscopic metallicity estimate of this object. Our estimates are [Fe/H]$_{\mathrm{K}}=0.26$ and [Fe/H]$_{\mathrm{H}}=0.31$ dex, supporting the findings of \cite{2011arXiv1111.2623A} and O11 that this star is metal-rich.

We compare our results to other spectroscopic studies in Table \ref{planettable}. Generally, our results are consistent within errors with those of both RA10 and O11. For the planets we have in common with RA10, our $K$-band estimate differences have mean~$=-0.11$ and $\sigma=0.08$ dex, and the $H$-band yields differences with mean $=-0.14$ and $\sigma = 0.12$ dex. For planets we have in common with O11, our $K$-band estimate differences have mean~$=0.04$ and $\sigma=0.09$ dex and the $H$-band differences have mean $=-0.01$ and $\sigma~=~0.10$ dex. The most discrepant points, the $H$-band estimates for GJ 849 and GJ 876, have high S/N which allowed us to probe frame-by-frame variability. The individual frames yielded highly variable (0.1-0.2\AA) EW measurements of the $H$-band \ion{K}{1} feature in both objects.

\subsection{Planet-Metallicity Relation\label{planetsection}}
Since we have obtained metallicity estimates for five M dwarfs with giant planets (of seven known) and four M dwarf planet hosts (of 12 known) without known giant planets, we can statistically asses whether M dwarfs with giant planets are metal-rich. We elect to use the nonparametric Mann-Whitney-Wilcoxon rank-sum test \citep{Mann:1947tc} on the averages of our $K$ and $H$-band metallicity estimates. This test determines whether one set of observations has preferentially higher values than another. For the samples of non-giant planet hosts ($n=4$, median $=-0.03$) and giant planet hosts ($n=5$, median $=0.27$), we find that the distributions differed significantly ($P < 0.01$). The $P$-value is the probability of obtaining a rank-sum at least as extreme as the one obtained under the null hypothesis that neither group has preferentially higher metallicities, and our alternative hypothesis was that the giant planet hosts were more metal-rich. Thus, for the set of M dwarf planet hosts for which we have consistent metallicity estimates, we find that giant planet hosts are indeed preferentially metal-rich compared to M dwarfs with less massive planets.

\section{Conclusion}
We obtained NIR spectra of 22 M dwarf companions to FGK stars with known metallicities. From measurements of the strengths of two $H$-band features and an H$_{2}$O index, we have derived a new empirical metallicity relation for M dwarfs in the $H$-band, parallel to that presented by RA10 in the $K$-band. This relation requires higher S/N spectra, but has smaller residuals. We used these calibrations to estimate metallicities for a subset of M dwarf planet hosts, producing the first empirically calibrated spectroscopic metallicity estimate for the Jupiter-host GJ 317 and finding it to be metal-rich. We also used a nonparametric test to show that M dwarfs with giant planets are preferentially metal-rich. Finally, we anticipate that the metallicity-sensitive $H$-band features used in our calibration will be critical starting points for metallicity analysis of high resolution $H$-band spectra of M dwarfs from large surveys like APOGEE.

\acknowledgments
We thank Kevin Luhman for obtaining observations. The authors are visiting astronomers at the Infrared Telescope Facility, operated by the University of Hawaii under Cooperative Agreement no.\ NNX-08AE38A with the National Aeronautics and Space Administration, Science Mission Directorate, Planetary Astronomy Program. We acknowledge support from the NAI and PSARC, and NSF grants AST-1006676 and AST-1126413.  The Center for Exoplanets and Habitable Worlds is supported by the Pennsylvania State University, the Eberly College of Science, and the Pennsylvania Space Grant Consortium. The authors acknowledge the significant cultural role and reverence that the summit of Mauna Kea has within the indigenous Hawaiian community. We thank the anonymous referee for valuable comments and suggestions.

\bibliographystyle{apj}

\begin{deluxetable}{lcccccrrlr}
\tabletypesize{\small}
\rotate{}
\tablecaption{M Dwarf Metallicity Calibrators\label{mcal_table}}
\tablewidth{0pt}
\tablehead{&\multicolumn{2}{c}{$K$-band} & \multicolumn{2}{c}{$H$-band} & & & & \multicolumn{2}{c}{Primary}  \\
\colhead{Name} & \colhead{EW$_{\mathrm{Na}}$[\AA]} & \colhead{EW$_{\mathrm{Ca}}$[\AA]} &\colhead{EW$_{\mathrm{Ca}}$[\AA]} &\colhead{EW$_{\mathrm{K}}$[\AA]} & \colhead{Est. SpT\tablenotemark{a}} & \colhead{[Fe/H]$_{\mathrm{K}}$} & \colhead{[Fe/H]$_{\mathrm{H}}$} & \colhead{Name} & \colhead{[Fe/H]$_{\mathrm{SPOCS}}$\tablenotemark{b}}}\startdata
HD 18143 C & 6.19$\pm$0.06 & 3.98$\pm$0.08 & 2.34$\pm$0.05 & 0.81$\pm$0.02 & M3.0 & $+0.17$ & $+0.09$ & HD 18143 & $+0.28$\\
J03480588+4032226\tablenotemark{b,c} & 7.59$\pm$0.07 & 5.50$\pm$0.09 & 3.51$\pm$0.05 & 0.91$\pm$0.02 & M2.0 & $+0.47$ & $+0.45$ & HD 23596 & $+0.22$\\
NLTT 14186 & 5.77$\pm$0.04 & 4.81$\pm$0.04 & 2.78$\pm$0.03 & 0.72$\pm$0.01 & M2.5 & $+0.17$ & $+0.20$ & HD 31412 & $+0.05$\\
GJ 3348 B & 5.12$\pm$0.10 & 3.15$\pm$0.06 & 1.59$\pm$0.03 & 0.94$\pm$0.02 & M3.5 & $-0.04$ & $-0.12$ & HD 35956 & $-0.22$\\
HD 38529 B & 6.84$\pm$0.07 & 5.11$\pm$0.09 & 3.45$\pm$0.05 & 0.71$\pm$0.02 & M0.0 & $+0.33$ & $+0.40$ & HD 38529 & $+0.45$\\
NLTT 16333 & 5.10$\pm$0.04 & 3.44$\pm$0.05 & 2.33$\pm$0.03 & 0.81$\pm$0.02 & M2.5 & $-0.03$ & $+0.01$ & GJ 9208 A & $-0.04$\\
HD 46375 B & 6.44$\pm$0.04 & 5.15$\pm$0.05 & 3.33$\pm$0.03 & 0.75$\pm$0.01 & M0.0 & $+0.28$ & $+0.34$ & HD 46375 & $+0.24$\\
GJ 250 B & 4.35$\pm$0.06 & 4.12$\pm$0.07 & 2.59$\pm$0.04 & 0.56$\pm$0.02 & M1.5 & $-0.08$ & $+0.03$ & GJ 250 A & $+0.14$\\
GJ 297.2 B & 4.56$\pm$0.10 & 3.82$\pm$0.11 & 2.56$\pm$0.08 & 0.45$\pm$0.03 & M1.5 & $-0.08$ & $+0.01$ & GJ 297.2 A & $-0.09$\\
GJ 324 B & 7.57$\pm$0.06 & 3.91$\pm$0.07 & 2.12$\pm$0.05 & 1.18$\pm$0.02 & M3.5 & $+0.36$ & $+0.16$ & GJ 324 A & $+0.31$\\
GJ 376 B & 6.86$\pm$0.05 & 1.95$\pm$0.07 & 1.38$\pm$0.05 & 2.02$\pm$0.02 & M6.5 & $+0.12$ & $+0.20$ & HD 86728 & $+0.20$\\
GJ 544 B & 5.14$\pm$0.06 & 2.66$\pm$0.07 & 1.44$\pm$0.04 & 0.94$\pm$0.02 & M4.0 & $-0.07$ & $-0.20$ & GJ 544 A & $-0.18$\\
GJ 611 B & 2.67$\pm$0.05 & 1.39$\pm$0.06 & 0.90$\pm$0.04 & 0.72$\pm$0.02 & M4.5 & $-0.51$ & $-0.41$ & HD 144579 & $-0.69$\\
NLTT 45791 & 4.63$\pm$0.04 & 3.94$\pm$0.04 & 1.97$\pm$0.03 & 0.47$\pm$0.01 & M1.5 & $-0.06$ & $-0.13$ & NLTT 45789 & $-0.06$\\
J19320809-1119573\tablenotemark{b,c} & 5.15$\pm$0.07 & 3.64$\pm$0.11 & 1.89$\pm$0.05 & 1.02$\pm$0.03 & M3.5 & $-0.00$ & $+0.00$ & HD 183870 & $+0.05$\\
GJ 768.1 B & 5.41$\pm$0.05 & 3.97$\pm$0.05 & 2.58$\pm$0.03 & 0.64$\pm$0.05 & M3.0 & $+0.06$ & $+0.08$ & GJ 768.1 A & $+0.16$\\
GJ 777 B & 5.25$\pm$0.04 & 3.00$\pm$0.04 & 2.63$\pm$0.03 & 0.70$\pm$0.02 & M4.5 & $-0.03$ & $+0.02$ & HD 190360 & $+0.21$\\
GJ 783.2 B & 4.01$\pm$0.04 & 2.65$\pm$0.04 & 1.77$\pm$0.02 & 0.70$\pm$0.01 & M4.5 & $-0.23$ & $-0.18$ & GJ 783.2 A & $-0.15$\\
GJ 797 B & 4.09$\pm$0.03 & 3.57$\pm$0.03 & 2.11$\pm$0.02 & 0.51$\pm$0.01 & M2.5 & $-0.16$ & $-0.13$ & GJ 797 A & $-0.09$\\
GJ 872 B & 3.77$\pm$0.04 & 2.96$\pm$0.06 & 1.50$\pm$0.03 & 0.43$\pm$0.02 & M2.0 & $-0.25$ & $-0.27$ & GJ 872 A & $-0.22$\\
LSPM J2335+3100 & 3.18$\pm$0.05 & 2.83$\pm$0.06 & 1.30$\pm$0.03 & 0.45$\pm$0.02 & M3.0 & $-0.34$ & $-0.38$ & HIP 116421 & $-0.40$\\
HD 222582 B & 5.10$\pm$0.08 & 3.43$\pm$0.11 & 1.96$\pm$0.06 & 0.84$\pm$0.03 & M3.5 & $-0.03$ & $-0.05$ & HD 222582 & $-0.03$\\
\enddata
\tablecomments{The list of M dwarf companions to higher mass primaries which are used as calibration targets for the metallicity relations.}
\tablenotetext{a}{Spectral types are estimated from the H$_{2}$O-K index value, following \cite{2010ApJ...722..971C}.}
\tablenotetext{b}{Primary metallicities from SPOCS, which adopts an uncertainty in [Fe/H] of 0.03 dex \citep{2005ApJS..159..141V}.}
\tablenotetext{c}{The 2MASS IDs are provided for these targets, which are new proper motion-selected binaries from \cite{2011arXiv1108.2719L}.}
\end{deluxetable}

\begin{deluxetable}{lccccc}
\tablecaption{M Dwarf Planet Hosts}
\tablewidth{0pt}
\tablehead{
\colhead{Name} & \colhead{[Fe/H]$_{\mathrm{K}}$} & \colhead{[Fe/H]$_{\mathrm{H}}$} & \colhead{RA10} & \colhead{O11} & \colhead{Known Giant Planet Host}
}
\startdata
GJ 436 & $-0.00$ & $+0.02$ & $-0.00$ & $+0.08\pm0.05$ & no \\
GJ 581 & $-0.04$ & $-0.09$ & $-0.02$ & $-0.15\pm0.03$ & no \\
HIP 79431 & $+0.53$ & $+0.50$ & $+0.60$ & $ \ldots $ & yes \\
GJ 649 & $-0.07$ & $-0.05$ & $+0.14$ & $ \ldots $ & no \\
HIP 57050 & $-0.03$ & $+0.05$ & $+0.12$ & $ \ldots $ & no \\
GJ 849 & $+0.31$ & $+0.22$ & $+0.49$ & $+0.35\pm0.10$ & yes \\
GJ 876 & $+0.26$ & $+0.11$ & $+0.43$ & $+0.12\pm0.15$ & yes \\
GJ 179 & $+0.20$ & $+0.18$ & $ \ldots $ & $ \ldots $ & yes \\
GJ 317 & $+0.26$ & $+0.31$ & $ \ldots $ & $+0.20\pm0.14$ & yes \\
\enddata
\tablecomments{Metallicity estimates for the observed planet hosts, along with estimates from O11 and RA10. Errors for both the H and K-band metallicity estimates are $\sigma=0.12$ dex.\label{planettable}}
\end{deluxetable}

\clearpage

\end{document}